\documentclass[
aps,%
12pt,%
final,%
notitlepage,%
oneside,%
onecolumn,%
nobibnotes,%
nofootinbib,% In the current version of REVTeX, when this option is on,
superscriptaddress,%
noshowpacs,%
centertags]%
{revtex4}
\RequirePackage{graphicx}

\newcommand{\Eq}[1]   {Eq.~(\ref{#1})}

\newcommand{\Fi}[1]   {Fig.~\ref{#1}}

\newcommand{\agev}    {\mbox{$A$~GeV}}               %PRL notation

\newcommand{\rb}[1]   {\mbox{\textrm{\small #1}}}
\newcommand{\rbt}[1]  {\mbox{\textrm{\tiny #1}}}
\newcommand{\sqrts}   {\ensuremath{\sqrt{s_{_{\rbt{NN}}}}}}

\newcommand{\lam}     {\ensuremath{\Lambda}}
\newcommand{\lab}     {\ensuremath{\bar{\Lambda}}}  
\newcommand{\xim}     {\ensuremath{\Xi^{-}}}

\newcommand{\pbar}    {\ensuremath{\bar{\textrm{p}}}}
\newcommand{\ommin}   {\ensuremath{\Omega^-}}
\newcommand{\omplus}  {\ensuremath{\bar{\Omega}^+}}           % with bar
\newcommand{\mtavg}   {\ensuremath{\langle m_{\rb{t}} \rangle - m_{\rb{0}}}}

\newcommand{\dndy}    {\ensuremath{\textrm{d}n/\textrm{d}y}}

\newcommand{\npart}   {\ensuremath{N_{\rb{part}}}}

\newcommand{\mub}     {\ensuremath{\mu_{\rbt{B}}}}

\def\refitem#1{\relax}

\begin{document}
\title{System Size Dependence of Particle Production at the SPS}

\author{\firstname{C.} \surname{Blume}}
\email{blume@ikf.uni-frankfurt.de}
\affiliation{Institut f\"ur Kernphysik, J.W.~Goethe-Universit\"at,  \\
             Max-von-Laue-Str.~1, D-60438 Frankfurt am Main, Germany}

\begin{abstract}
Recent results on the system size dependence of net-baryon and hyperon
production as measured at the CERN SPS are discussed.  The observed
\npart~dependences of yields, but also of dynamical properties, such
as average transverse momenta, can be described in the context of the
core corona approach.  Other observables, such as antiproton yields
and net-protons at forward rapidities, do not follow the predictions
of this model.  Possible implications for a search for a critical
point in the QCD phase diagram are discussed.  Event-by-event
fluctuations of the relative core to corona source contributions might
influence fluctuation observables (e.g. multiplicity fluctuations).
The magnitude of this effect is investigated.
\end{abstract}

\maketitle

\section{Introduction}

The system size dependence of particle production in heavy ion
collisions can be measured in two ways: either by comparing central
reactions of nuclei of different size, or by analyzing centrality
selected minimum bias collisions of nuclei of fixed size.  It turns
out that many observables follow a very similar \npart~dependence, if their
centrality dependence is studied.  Usually a rapid change for very
peripheral reactions is followed by only a weak variation towards very
central collisions.  This indicates that many features of system size
dependences are dominated by the same underlying geometrically
effects.  It is therefore important to find a reliable way of modeling
them in order to extract non-trivial effects.

\section{Core Corona Separation}

A quite successful way of describing the main features of system size
dependences of particle production is the core corona approach
\cite{BOZEK,BECATTINI1,BECATTINI2,AICHELIN1}.  In this model a 
nucleus-nucleus collision is decomposed into a central core, which 
corresponds to the large fireball produced in central A+A collisions,
and a surrounding corona, which is equivalent to independent
nucleon-nucleon reactions.  To quantify the relative contribution of
the two components the fraction of nucleons that scatter more than
once, $f(\npart)$, can be used.  $f(\npart)$ can simply be calculated
within a Glauber model \cite{GLAUBER,REYGERS}.  This quantity allows
for a natural interpolation between the yields $Y$ measured in
elementary p+p ($= Y_{\textrm{corona}}$) and in central nucleus-nucleus
collisions ($= Y_{\textrm{core}}$):
\begin{equation}
\label{eq:corecorona}
Y(\npart)  =  \npart \: [ f(\npart) \: Y_{\textrm{core}}  \: 
                        + \: (1 - f(\npart)) \: Y_{\textrm{corona}} ]
\end{equation}

The left panel of \Fi{fig:glauber} shows $f(\npart)$ for various
symmetric reaction systems.  Due to their different surface to volume
ratio, the \npart\ dependence of $f(\npart)$ is much steeper for
smaller reaction systems than for larger ones.  This feature of the
model allows, for instance, to describe the centrality dependences of
strange particle production measured for Cu+Cu and Au+Au collisions at
RHIC \cite{TIMMINS}.  An important feature of the curves shown in the
left panel of \Fi{fig:glauber} is that the maximum value of
$f(\npart)$ depends on the size of the nuclei.  While for very central
Pb+Pb collisions $f_{\rb{max}} \approx 0.9$ can be reached, the
maximum value for very central C+C reactions is significantly lower:
$f_{\rb{max}} \approx 0.65$.  This $A$ dependence of $f_{\rb{max}}$ is
illustrated by the dashed red line.  It has the important consequence
that the different relative core corona contributions have to be taken
into account when comparing even very central nucleus-nucleus
collisions of different size, which usually are considered to
represent only the core-like fireball.  This effect might explain any
observed system size dependence of the chemical freeze-out parameters
$T$ and \mub, as derived from statistical model fits to central A+A
collisions for nuclei of different size (e.g. \cite{BECATTINI3}).  As
a consequence of this it therefore follows that the system size does
not provide a good control parameter to probe different regions of the
QCD phase diagram via the production of fireballs of different
temperature \cite{BLUME1}.  It is thus more likely that one only
observes a change in the relative admixture of central fireball (core)
and peripheral p+p like corona, whose freeze-out parameters are
different, but independent of the size of the involved nuclei.  In any
case the underlying effect from the core corona separation should be
taken into account when chemical freeze-out parameters are extracted.

Another interesting aspect of the core corona approach is shown in the
right panel of \Fi{fig:glauber}.  Here, the \npart\ dependence of
$f(\npart)$ is shown for asymmetric collisions.  In this case quite
distinct centrality dependences can be observed.  While for symmetric
collisions this dependence is following a continuous rise (left panel
of \Fi{fig:glauber}), for asymmetric collision systems with a small
projectile nucleus (e.g. O+Pb or Si+Pb) a rapid rise followed by a
maximum and a subsequent decrease of $f(\npart)$ is seen.  Therefore,
a similar centrality dependence of particle yields can be expected for
these type of collisions and its measurement would constitute a test
for the validity of the core-corona model \cite{BLUME1}.

\section{Strangeness Production}

The left panel of \Fi{fig:meanmt_vs_nw} shows the system size
dependence of the strangeness enhancement factors $E$ measured by the
NA49 collaboration at \sqrts~= 17.3~GeV \cite{NA49HYSDEP,BLUME2}.  For
this measurement $E$ is defined relative to p+p reactions as baseline:
\begin{equation}
E = \left. \left( \frac{1}{\langle \npart \rangle} 
           \left. \frac{\textrm{d}N(\textrm{Pb+Pb})}{\textrm{d}y}\right|_{y=0}
    \right) \right/
    \left( \frac{1}{2} 
           \left. \frac{\textrm{d}N(\textrm{p+p})}  {\textrm{d}y}\right|_{y=0}
    \right)
\end{equation}
As a general feature a rapid rise of $E$ for $\npart < 60$ is
seen, which then turns into a slow increase with \npart\ for \lam,
\xim, and $\Omega$, or even into a saturation (\lab).  Comparable
observations have been made by the NA57 experiment \cite{NA57ENHANCE}
at the same \sqrts\ and by the STAR collaboration at \sqrts~= 200~GeV
\cite{STARENHANCE}.

In the context of statistical models, the system size dependence of
strange particle production was expected to be described by the
transition from a canonical to a grand canonical ensemble
\cite{HAMIEH}.  This effect provides a natural explanation of the
hierarchy of the suppression pattern, which depends on the strangeness
content of the particles ($E(\lam) < E(\Xi) < E(\Omega)$).  However,
the onset of the enhancement is predicted to happen for smaller
systems ($\npart \approx 30$), than observed in the data.  Also, this
model does not explain the slow rise with \npart\ seen for $\Xi$ and
$\Omega$ for mid-central and central collisions.  However,  within the
core corona model this behavior can be described naturally, as
described in the previous section.  This is illustrated by the solid
lines shown in the left panel of \Fi{fig:meanmt_vs_nw}.

It is interesting to note that this approach not only matches the
observed centrality dependences of yields, but does also describe
dynamical quantities such as the average transverse momenta \mtavg.
This is shown in the right panel of \Fi{fig:meanmt_vs_nw} for strange
particles and (anti-)protons \cite{NA49HYSDEP,NA49SDPR1}.  A similar
observation has been made for elliptic flow at RHIC \cite{AICHELIN2}.

\section{Centrality Dependence of Net Protons}

Recent data on proton and antiproton production allow to study the
system size dependence of stopping \cite{NA49SDPR2}.  In this analysis
the rapidity spectra of net-protons ($\textrm{p}-\pbar$) have been
determined in several bins of centrality for minimum bias Pb+Pb
collisions at 40$A$ and 158\agev.  Based on this data the centrality
evolution of net-proton yields at different rapidities can be
investigated.  The left panel of \Fi{fig:net_protons_2} shows a
comparison of the net-proton yields normalized by \npart\ ($1/\npart \:
\dndy(\textrm{p}-\pbar)$) as measured at midrapidity and at forward
rapidity at 158\agev.  While the midrapidity net-proton yields are
slightly rising from peripheral to central Pb+Pb collisions, the
forward yields exhibit a significantly different centrality
dependence.  Here the normalized yields are constant from very
peripheral on or even drop towards very central collisions.  In
contrast to the net-protons at midrapidity, the forward yields can
therefore also not be described by the core corona approach.  This
implies that for the system size dependence of stopping, i.e. the
processes that transport baryon number along the longitudinal axis and
which dominate the net-proton spectra at forward rapidities, the
distinction between core and corona is not important.  These processes
do not seem to depend to strongly on the number of collisions that a
single nucleon experiences.

For the yield of antiprotons a different centrality dependence is
observed than for other particles \cite{NA49SDPR2} (right panel of
\Fi{fig:net_protons_2}).  In this case the yield per participant is
decreasing from p+p collisions towards central Pb+Pb reactions.  This
is most likely caused by absorption antibaryons in the fireball
medium.  Also this effect is difficult to describe in the core corona
framework, since it rather depends on the average path length of the
antiprotons inside the fireball medium than on relative size of core
and corona.  The resulting antiproton yields will thus be rather a
complicated interplay between the system size dependences of
production and absorption mechanism.

\section{Multiplicity Fluctuations}

The system size dependence of multiplicity fluctuations was measured
by the NA49 experiment \cite{NA49SDMFLUC}.  The fluctuations,
quantified by the variance of the event-by-event measured multiplicity
$n$ normalized by its mean $\omega = \textrm{Var}(n)/\langle n
\rangle$, exhibit a quite distinct centrality dependence.  While at
\sqrts~= 17.3~GeV $\omega \approx 1$ is found for p+p collisions, very
peripheral Pb+Pb reactions result in much higher fluctuations ($\omega
\approx 3$ for all charged particles) which then decrease towards very
central collisions where $\omega$ close to unity is observed again.
In the following we investigate to what extend these fluctuations
could be caused by geometry fluctuations that are intrinsic to the
core corona approach.  This source of fluctuations is of special
interest, since it will be present even in the idealized situation
of a perfect centrality selection, corresponding to the hypothetical
situation where the number of participants could be fixed exactly in
the experiment.  Also in the case \npart~=~fixed, the relative
contribution of the core and the corona, as given by $f(\npart)$, will
still fluctuate from event to event.  The upper left panel of
\Fi{fig:ffluct} shows the distribution of $f$ versus \npart, as
calculated within a Glauber model \cite{GLAUBER,REYGERS}.  The
event-by-event distribution of $f$ is much wider for peripheral
reactions than for central ones.  The resulting variance
$\textrm{Var}(f)$ (see upper right panel of \Fi{fig:ffluct}) has a
\npart\ dependence that is quite similar to the one observed for
multiplicity fluctuations.  Since the particle yield per participating
nucleon is smaller for the corona than for the core part, the
fluctuations in $f$ will translate into multiplicity fluctuations that
should be part of the observed ones.  To evaluate how large this
contribution can be, a simple Monte Carlo model was constructed.  In
this model the multiplicity of a given event is calculated according
to \Eq{eq:corecorona}.  The event-by-event yields of the core and
corona sources, $Y_{\textrm{core}}$ and $Y_{\textrm{corona}}$, are
generated from Poisson distributions, whose means have been adjusted
to the measured midrapidity yields per participating nucleon of
charged pions in central Pb+Pb (core) \cite{NA49PIPBPB} and p+p
(corona) \cite{NA49PIPP} collisions.  The value of $f$ for a given
\npart\ is sampled from the distribution shown in \Fi{fig:ffluct}.  In
the lower left panel of \Fi{fig:ffluct} the result of this simulation
is shown.  This way one obtains multiplicity fluctuations whose
centrality dependence has the same shape as the one observed in the
data.  However, the magnitude of the fluctuations is much higher in
the data than what results from this model study (see the comparison
in the lower right panel of \Fi{fig:ffluct}).  The difference between
the yields $Y_{\textrm{core}}$ and $Y_{\textrm{corona}}$ would need to be
two orders of magnitude higher to reach the level of the measurement.

\section{Conclusions}

Many observables measured in heavy ion reactions turn out to follow
a very similar centrality dependence, which is characterized by a rapid
rise in the region of very peripheral collisions followed by a
significantly slower increase towards very central ones.  This include
the yields per participant of strange particles, but also their mean
transverse mass and even elliptic flow $v_{2}$.  It is found that
the core corona model provides a very effective way to describe
the main features of this geometry driven behavior and thus
constitutes a good comparison baseline which allows to separate
non-trivial effects.  Observed exceptions to this general picture are,
for instance, the centrality dependence of antiproton yields and
net-proton \dndy\ at forward rapidities, where more complicated
physical processes are involved (absorption of antibaryons, baryon
number transport) which cannot easily be described by the core corona
approach.  For asymmetric collisions a significantly different
centrality dependence than for symmetric ones is predicted by this
model, which is characterized by a maximum for mid-central and a drop
towards central collisions.  Therefore the investigation of asymmetric
collisions might provide a good test for the core corona approach.
Another consequence of this model is that the observed change of the
chemical freeze-out temperature as derived from statistical model
analyses is basically only due to a change of the relative admixture
of the core and corona contributions.  Therefore, the system size
might not provide a good control parameter to scan the QCD phase
diagram, since the properties of the central core fireball might in
fact not change and the system always follows the same trajectories
through the phase diagram.  Event-by-event fluctuations of the
relative core and corona contribution, which must be present even in
the case of an ideal centrality selection (i.e. \npart~=~fixed), do
contribute to multiplicity fluctuations.  However, the magnitude of
this effect is way too small to explain the measured data.  Still,
further studies might help to understand the system size dependences
of fluctuations of non-intensive quantities, such as
K/$\pi$~fluctuations.

\newpage

\begin{acknowledgments}
The author would like to thank K.~Reygers for the help with the Glauber
model calculations and acknowledges many stimulating and helpful
discussions with J.~Aichelin and F.~Becattini.
\end{acknowledgments}

\newpage

%----------------------------------------------------------------------
%
\begin{figure}[ht]
\begin{center}
\begin{minipage}[b]{0.49\linewidth}
\begin{center}
\includegraphics[width=\linewidth]{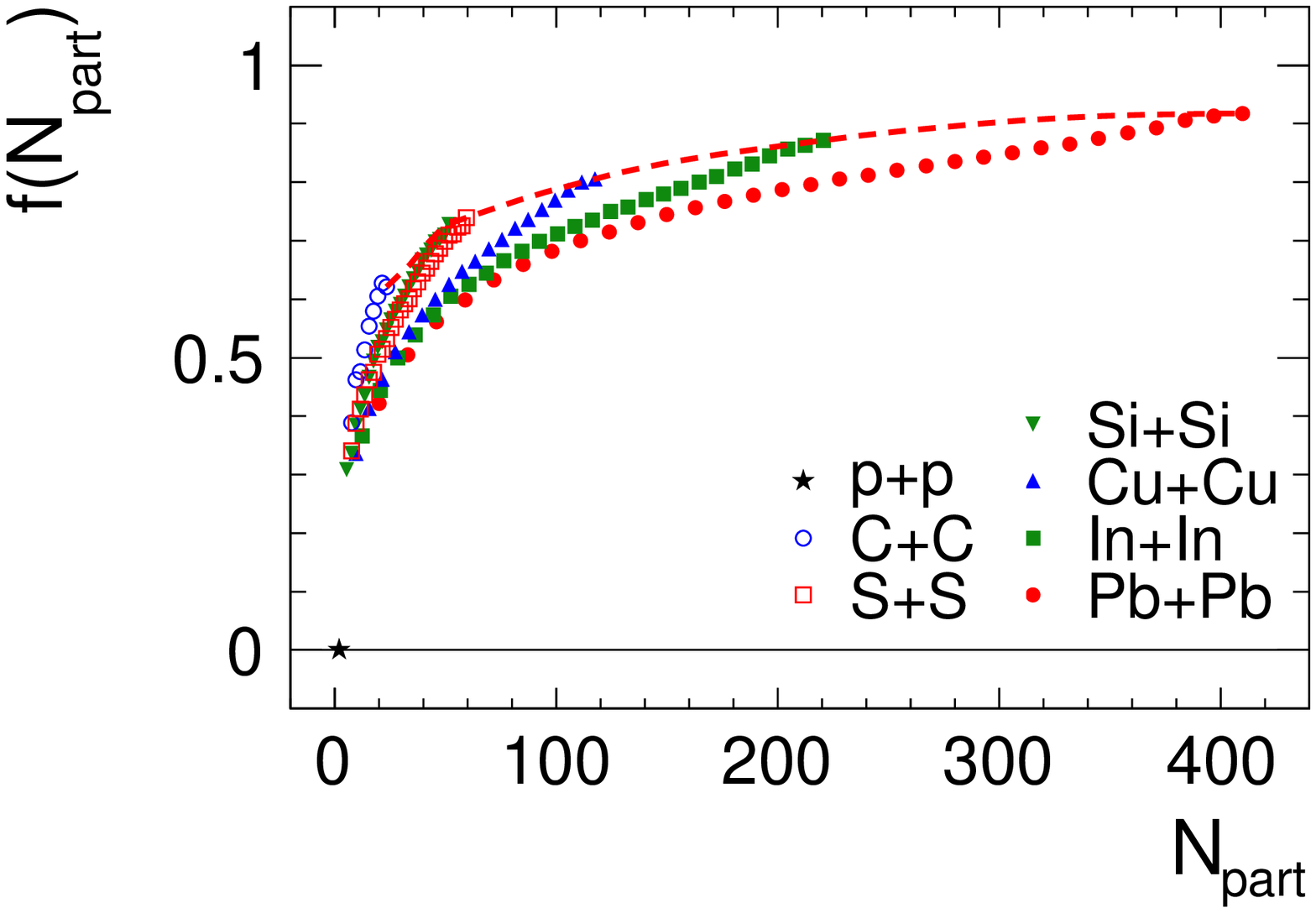}
\end{center}
\end{minipage}
\begin{minipage}[b]{0.49\linewidth}
\begin{center}
\includegraphics[width=\linewidth]{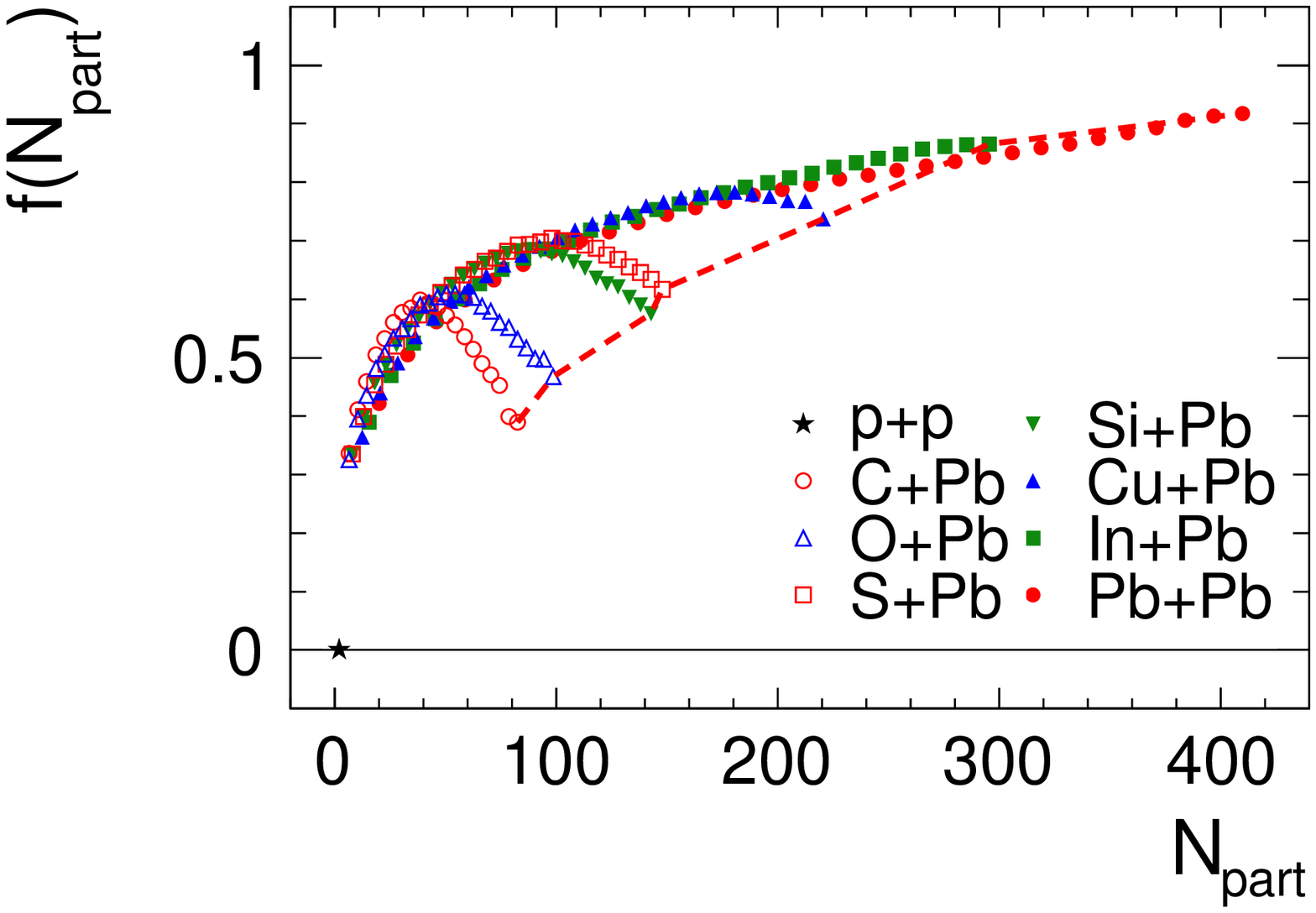}
\end{center}
\end{minipage}
\end{center}
\caption{}
\label{fig:glauber} 
\end{figure}
%
%----------------------------------------------------------------------

%----------------------------------------------------------------------
%
\begin{figure}[ht]
\begin{center}
\begin{minipage}[b]{0.42\linewidth}
\begin{center}
\includegraphics[width=1.00\linewidth]{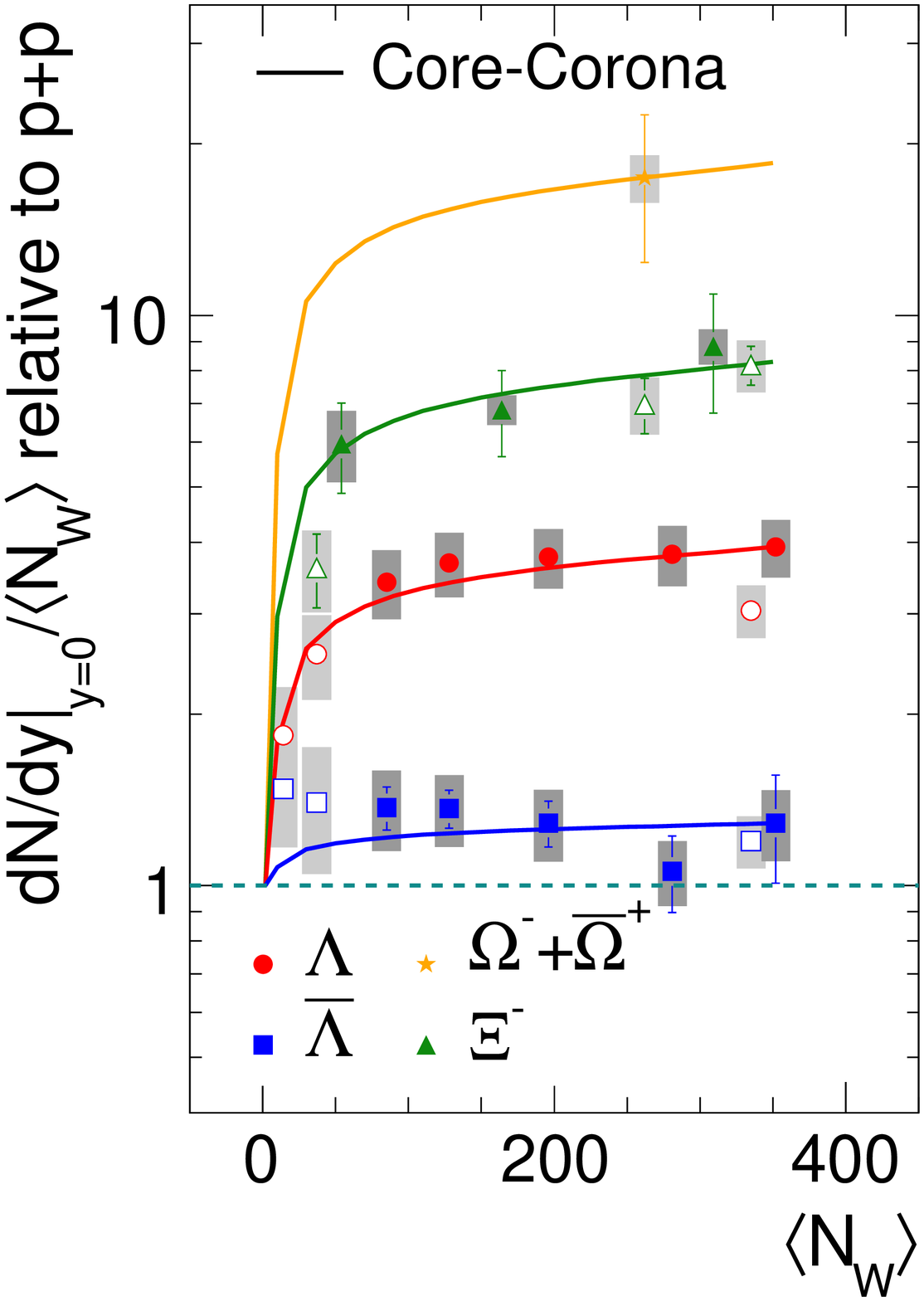}
\end{center}
\end{minipage}
\hspace{0.07\linewidth}
\begin{minipage}[b]{0.42\linewidth}
\begin{center}
\includegraphics[width=0.96\linewidth]{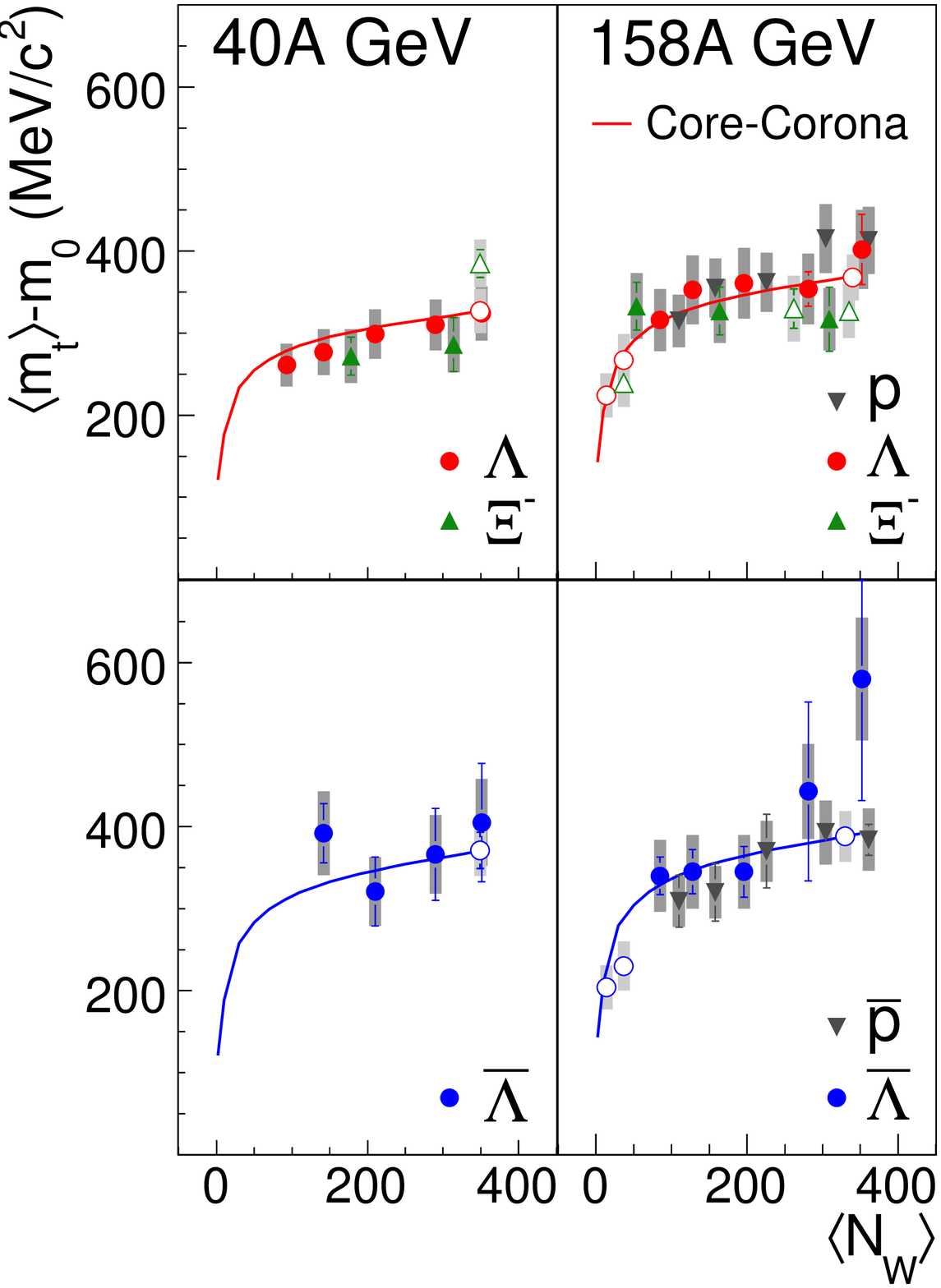}
\vspace{0.45cm}
\end{center}
\end{minipage}
\end{center}
\caption{}
\label{fig:meanmt_vs_nw} 
\end{figure} 
%
%----------------------------------------------------------------------

%----------------------------------------------------------------------
%
\begin{figure}[ht]
\begin{center}
\begin{minipage}[b]{0.47\linewidth}
\begin{center}
\includegraphics[width=1.00\linewidth]{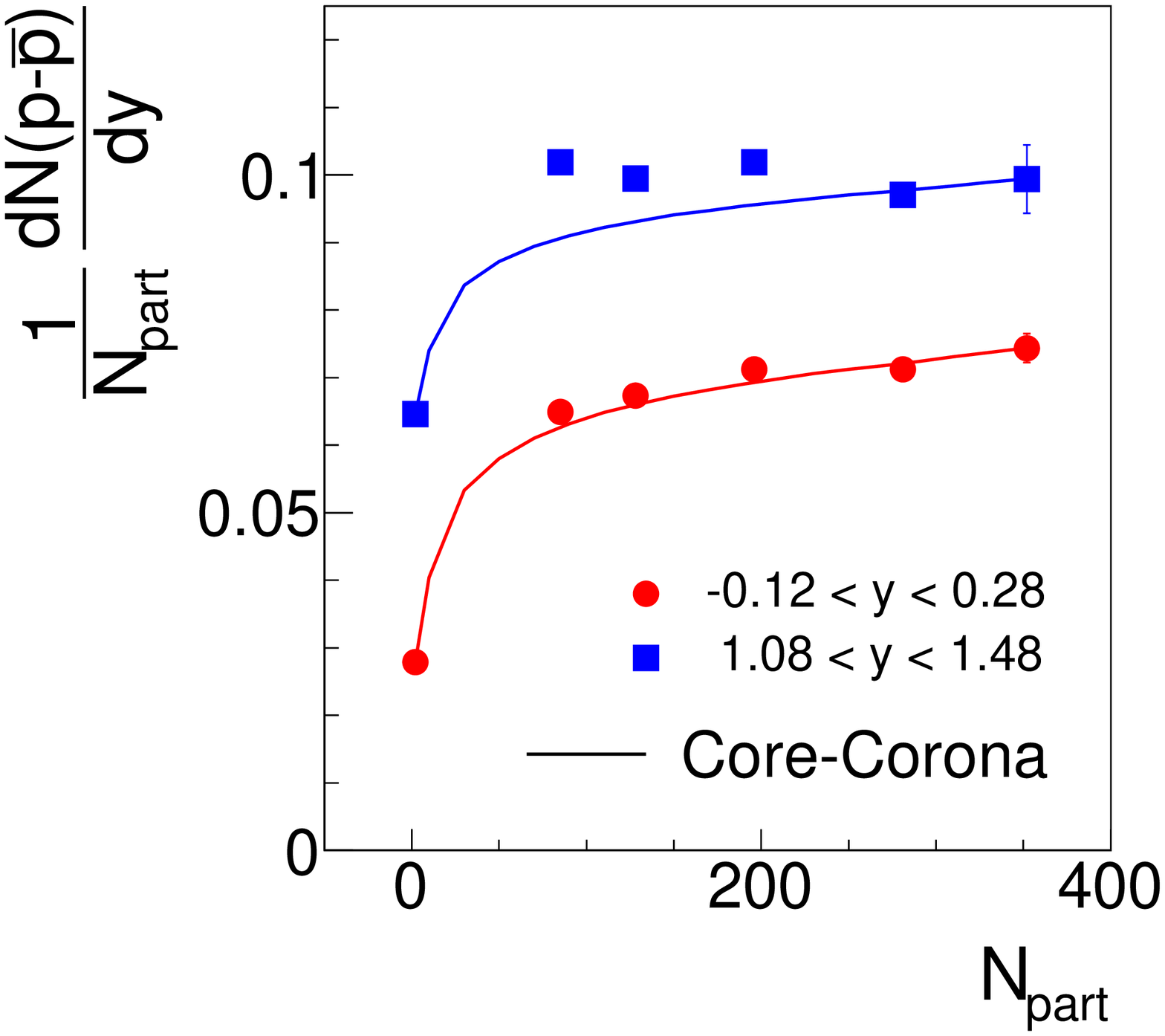}
\end{center}
\end{minipage}
\hspace{0.05\linewidth}
\begin{minipage}[b]{0.45\linewidth}
\begin{center}
\includegraphics[width=1.00\linewidth]{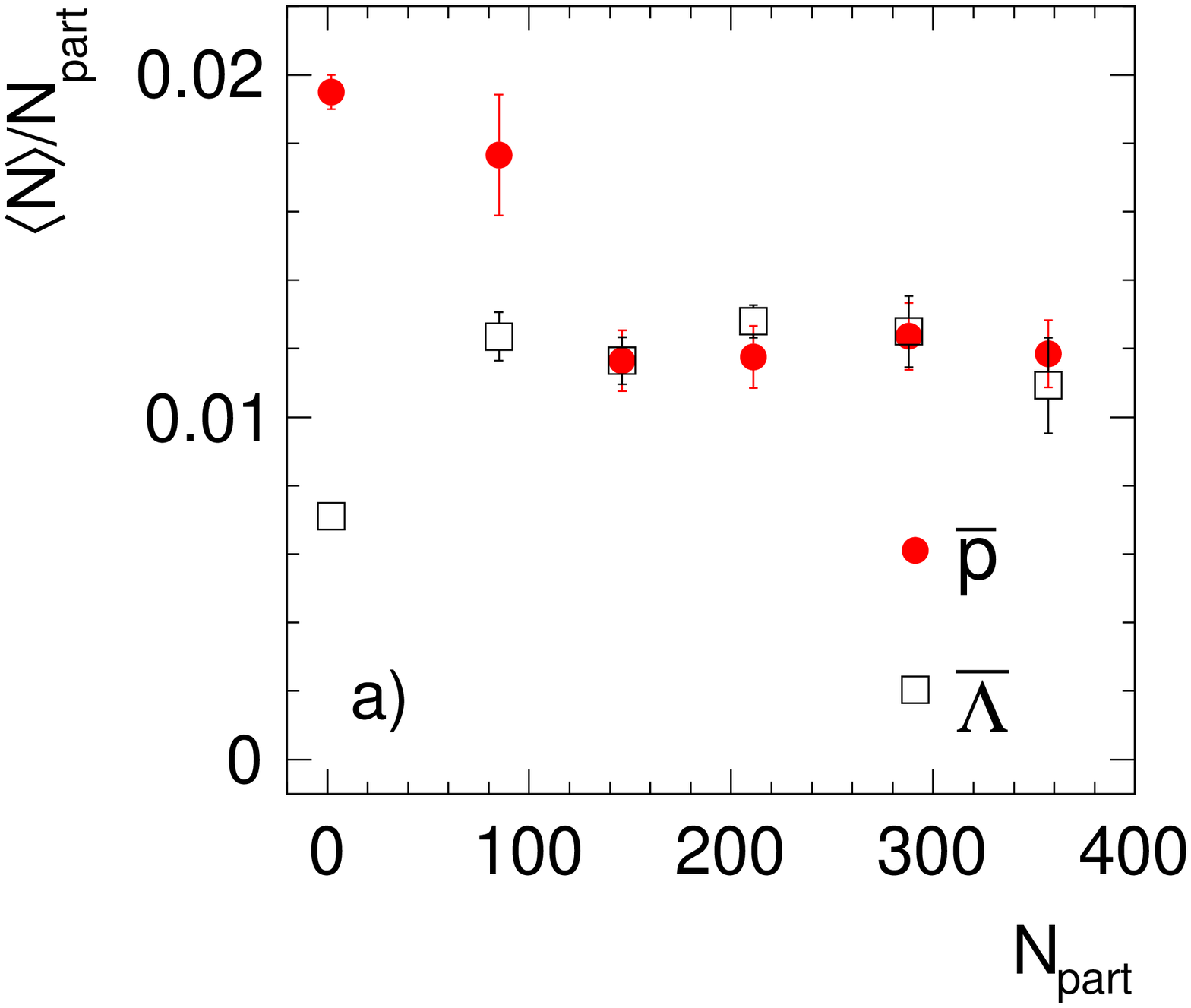}
\end{center}
\end{minipage}
\end{center}
\caption{}
\label{fig:net_protons_2} 
\end{figure} 
%
%----------------------------------------------------------------------

%----------------------------------------------------------------------
%
\begin{figure}[ht]
\begin{center}
\begin{minipage}[b]{0.49\linewidth}
\begin{center}
\includegraphics[width=\linewidth]{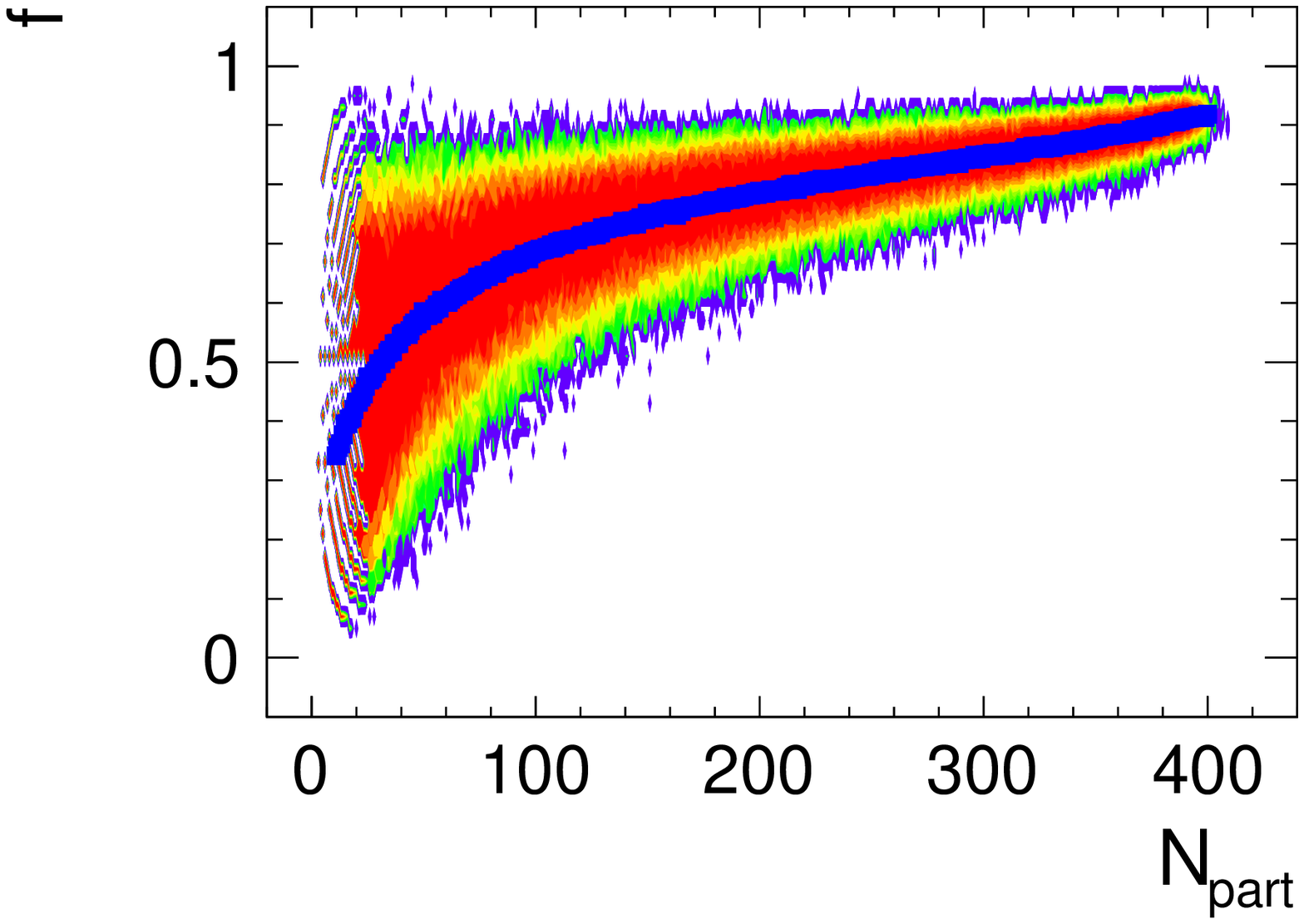}
\end{center}
\end{minipage}
\begin{minipage}[b]{0.49\linewidth}
\begin{center}
\includegraphics[width=\linewidth]{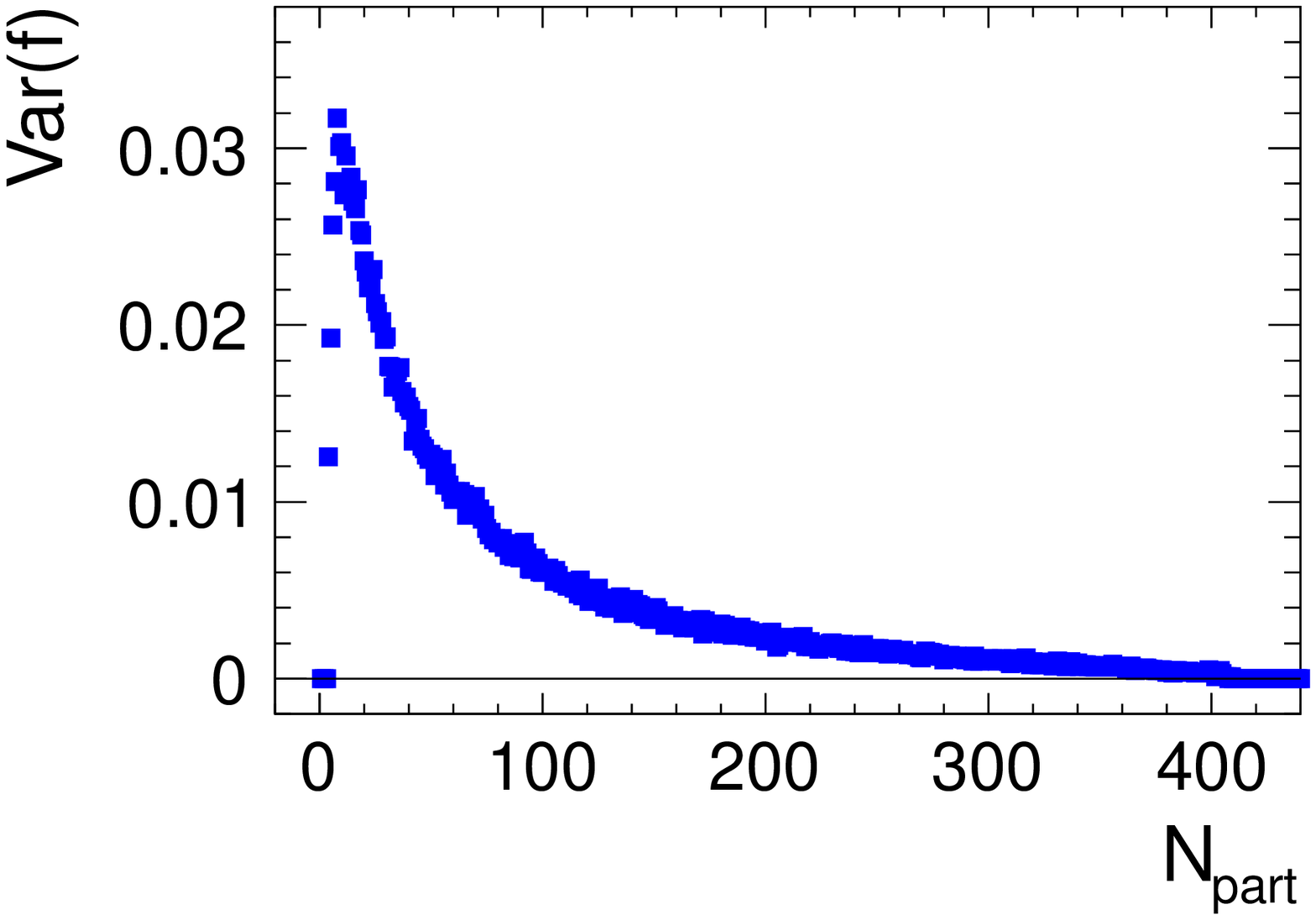}
\end{center}
\end{minipage}
\begin{minipage}[b]{0.49\linewidth}
\begin{center}
\includegraphics[width=\linewidth]{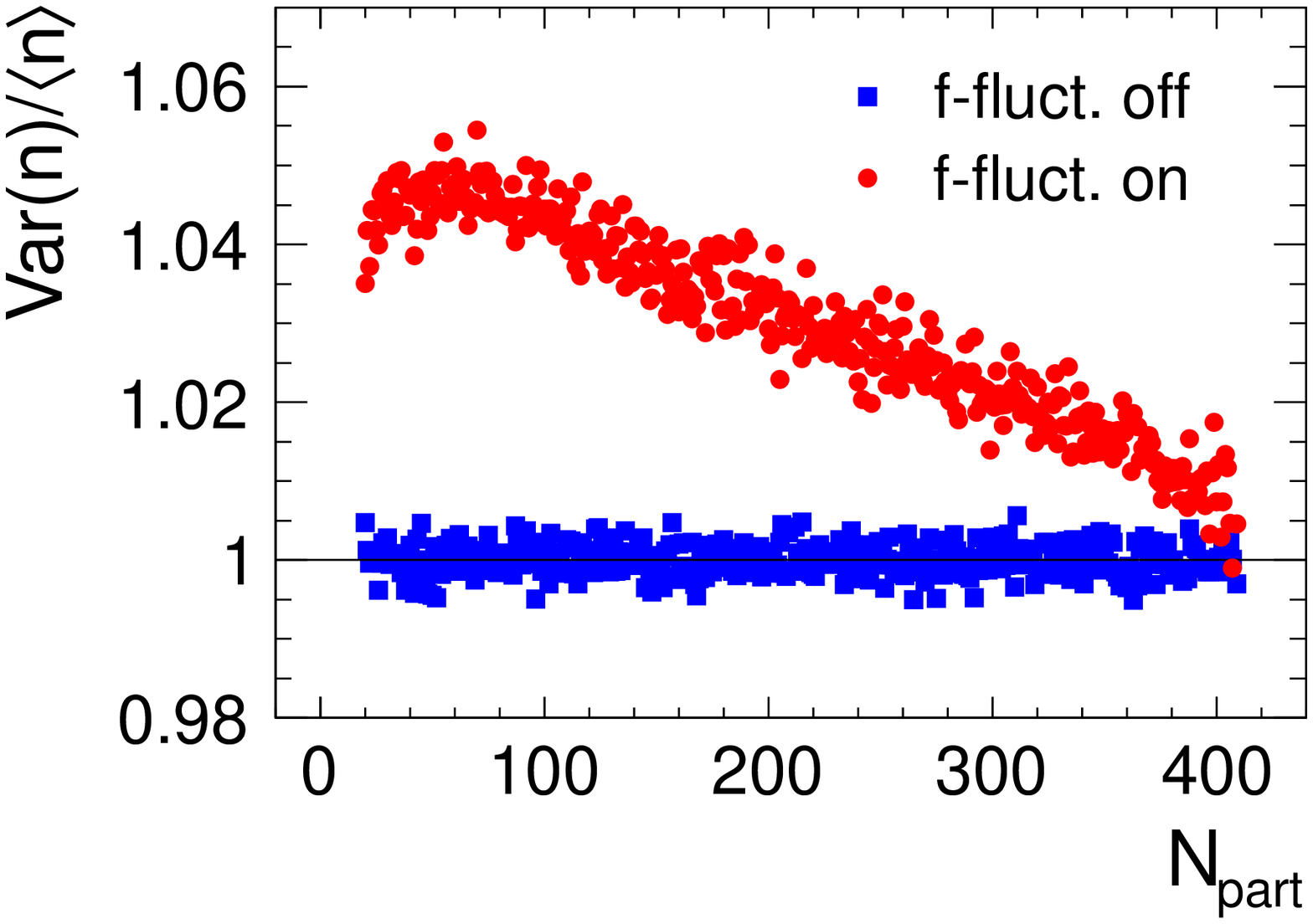}
\end{center}
\end{minipage}
\begin{minipage}[b]{0.49\linewidth}
\begin{center}
\includegraphics[width=\linewidth]{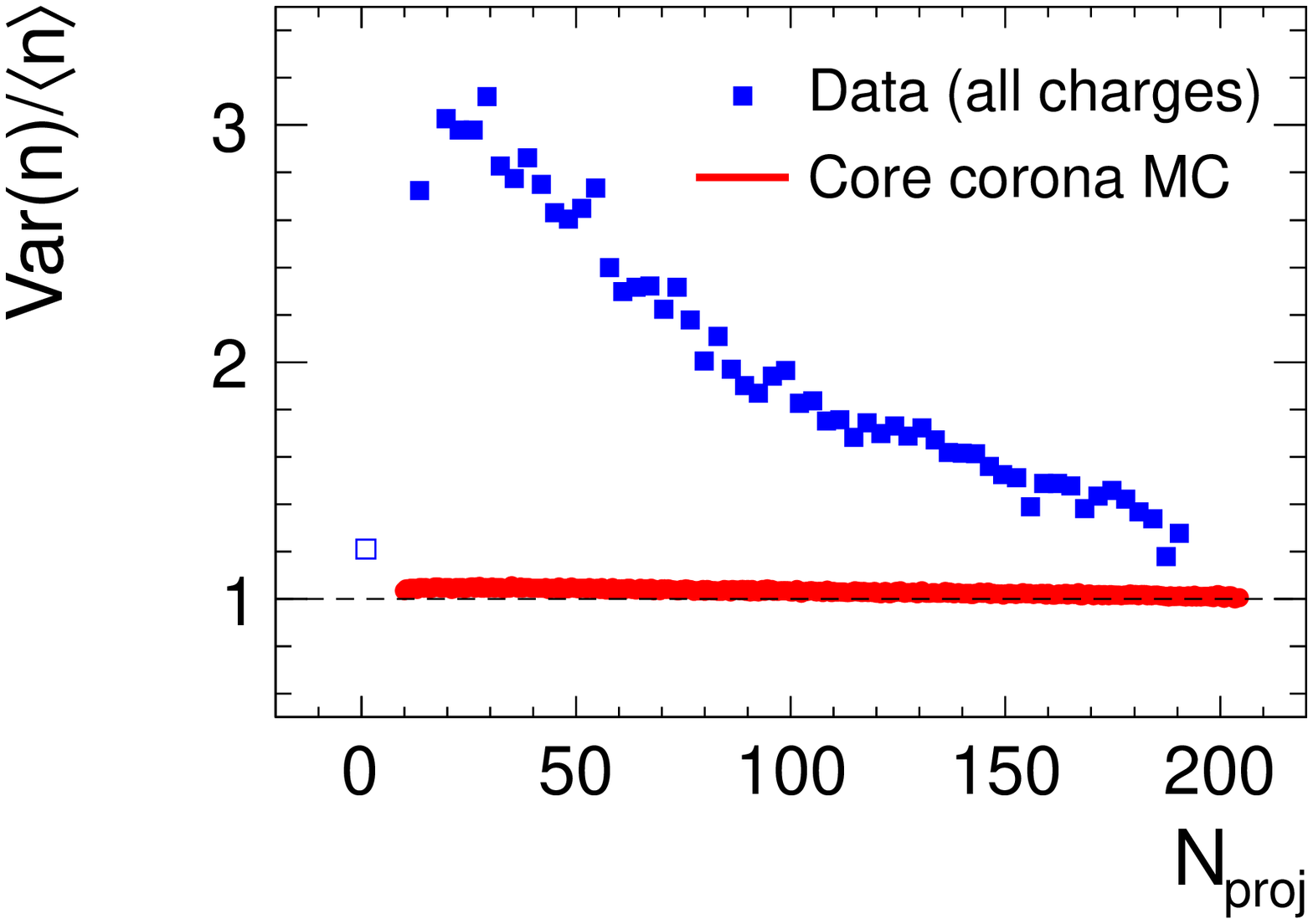}
\end{center}
\end{minipage}
\end{center}
\caption{}
\label{fig:ffluct} 
\end{figure}
%
%----------------------------------------------------------------------

\clearpage

\begin{center}
FIGURE CAPTIONS
\end{center}

\begin{enumerate}

\item[FIG. 1:] The fraction $f(\npart)$ of participating nucleons that
      scatter more than once as a function of the number of
      participants \npart.  $f(\npart)$ was calculated within a Glauber
      model \cite{GLAUBER,REYGERS}.  The left panel shows results for
      symmetric systems, the right panel for asymmetric ones.  The
      dashed lines connect the values for the most central collisions.

\item[FIG. 2:] Left: The midrapidity yields of \lam, \lab, \xim, and
      $\ommin+\omplus$ per wounded nucleon relative to p+p
      yields for central C+C, Si+Si and minimum bias Pb+Pb reactions
      at 158\agev\ \cite{NA49HYSDEP,BLUME2}.
      Right: The \mtavg\ values of strange particles and
      (anti-)protons at mid-rapidity for Pb+Pb collisions
      at 40$A$ and 158\agev, as well as for near-central C+C and Si+Si
      reactions at 158\agev\ \cite{NA49HYSDEP}.  The (anti-)proton
      data are taken from \cite{NA49SDPR1}.  Also shown are the
      results from a fit for \lam\ and  \lab\ with the core-corona
      approach (solid lines).

\item[FIG. 3:] Left: The net-proton yields per number of wounded
      nucleons at forward rapidity and at midrapidity for centrality
      selected Pb+Pb collisions at 158\agev.
      Right: The total antiproton and antilambda yields per number
      of wounded nucleon at midrapidity for centrality selected Pb+Pb
      collisions at 158\agev\ \cite{NA49SDPR2}.
 
\item[FIG. 4:] The distribution of the fraction of nucleons that
      scatter more than once $f(\npart)$ (upper left) and the
      corresponding variance $\textrm{Var}(f)$ (upper right), both as
      a function of \npart\ as calculated for Pb+Pb collisions at
      158\agev.  The lower left panel shows the multiplicity
      fluctuations generated with the MC model (see text for details).
      The red circles display the result if event-by-event fluctuations
      of $f$ are included, while the blue boxes correspond to the case
      where they are switched off.  The lower right panel shows
      multiplicity fluctuations for all charges versus the number of
      projectile participants as measured in Pb+Pb collisions at
      158\agev\ \cite{NA49SDMFLUC} (blue boxes).  The red line is the
      same MC result as shown in the left panel.

\end{enumerate}

\end{document}